# 3D Gamma-ray and Neutron Mapping in Real-Time with the Localization and Mapping Platform from Unmanned Aerial Systems and Man-Portable Configurations

R. Pavlovsky[1], J. W. Cates[1], W. J. Vanderlip[2], T. H.Y. Joshi[1], A. Haefner[1], E. Suzuki[1], R. Barnowski[2], V. Negut[1], A. Moran[1], K. Vetter[1,2], and B.J. Quiter[1]

[1]Applied Nuclear Physics, Lawrence Berkeley National Laboratory, Berkeley CA
[2]Department of Nuclear Engineering, University of California Berkeley, Berkeley CA

**Abstract**
Nuclear Scene Data Fusion (SDF), implemented in the Localization and Mapping Platform (LAMP) fuses three-dimensional (3D), real-time volumetric reconstructions of radiation sources with contextual information (e.g. LIDAR, camera, etc.) derived from the environment around the detector system. This information, particularly when obtained in real time, may be transformative for applications, including directed search for lost or stolen sources, consequence management after the release of radioactive materials, or contamination avoidance in security-related or emergency response scenarios. 3D reconstructions enabled by SDF localize contamination or hotspots to specific areas or objects, providing higher resolution over larger areas than conventional 2D approaches, and enabling more efficient planning and response, particularly in complex 3D environments.

In this work, we present the expansion of these gamma-ray mapping concepts to neutron source localization. This is achieved by integrating LAMP with a custom $Cs_2LiLa(Br,Cl)_6$:Ce (CLLBC) scintillator detector sensitive to both gamma rays and neutrons, which we refer to as Neutron Gamma LAMP (NG-LAMP). NG-LAMP enables simultaneous neutron and gamma-ray mapping with high resolution gamma-ray spectroscopy. The resolution of the system is about 3.5% Full Width At Half Maximum (FWHM) at 661.7 keV. With NG-LAMP and the SDF architecture, we demonstrate the ability to detect and localize surrogate Special Nuclear Materials (SNM) in real-time and in 3D based on neutron signatures alone, which is critical for the detection of heavily shielded SNM when gamma-ray signatures are attenuated. In this work, we show for the first time the ability to localize a neutron source in the presence of a strong gamma-ray source, the ability to simultaneously and spectroscopically localize three gamma-ray sources and a neutron source, and finally we present the localization of a surrogate SNM source based on neutron signatures alone, where gamma-ray emissions are shielded such that the gamma-dose rate at the exterior of the shielding is indistinguishable from background. In this final scenario we demonstrate that the gamma-ray background, reconstructed with the same parameters as the neutron reconstruction, is a near uniform distribution as one might expect in the case of heavily attenuated gamma-ray signal.

Introduction

Three-dimensional (3D) nuclear Scene Data Fusion (SDF), developed at Lawrence Berkeley National Laboratory [1,2], enables the rapid localization, identification, and visualization of point and distributed radioactive sources in 3D and real-time [2]. SDF produces rich data products with contextual information about 3D scenes fused with radiation data to provide radiation maps of objects or areas of interest. The potential applications of this technology are broad, and benefit scenarios that include directed search for lost or stolen sources, consequence management after the release of radioactive materials, or for contamination avoidance in security or emergency response scenarios [3]. These applications present many challenges due to the complexity of the scenes, the need for real-time situational awareness and easily communicable data products. SDF is uniquely capable of addressing these challenges, particularly when deployed via mobile platforms, including on small Unmanned Aerial Systems (sUAS), which can provide a safer way to map large, complex areas that are tens of thousands of square meters in size, with increased accuracy over conventional, static detection systems while producing detailed and feature-rich 3D data products.

We have reported on the most recent implementation of the SDF architecture, the compact, lightweight, multi-sensor Localization and Mapping Platform (LAMP) for detection and localization of gamma-ray point and distributed sources [1]. These previous demonstrations highlight data products in a wide variety of environments and scenarios for applications mentioned above. In this work, we extend the concept of 3D, real-time gamma-ray mapping to enable simultaneous gamma-ray and neutron mapping in 3D and in real-time. This is accomplished with four monolithic radiation detectors sensitive to both gamma-ray and neutron sources, fused by SDF in Neutron Gamma LAMP (NG-LAMP). NG-LAMP consists of an onboard computer, power distribution, and a Radiation Monitoring Devices Inc. (RMD) CLLBC scintillator [4-6] coupled to an LBNL custom electronics package, totaling 130 cm$^3$ of detector material with gamma-ray energy resolution of ~3.5% full-width at half-maximum (FWHM) at 661.7 keV. NG-LAMP produces 3D data products while remotely streaming compressed 2D maps to the user in real-time in handheld mode or on unmanned ground or aerial platforms. In this work, we show results from the first demonstration of 3D, real-time simultaneous neutron and gamma-ray mapping from a sUAS and from handheld deployments. We also present the first demonstration of real-time multiple source reconstructions of both gamma-rays and neutrons from handheld measurements, as well as real-time simultaneous neutron and gamma-ray mapping in 3D for a shielded surrogate of ~1 significant quantity (SQ) [10] of Pu, for a neutron point source in gamma-ray background, and a neutron point source in the presence of a strong gamma-ray source.

**Methods and Approach**

NG-LAMP is one implementation of LAMP, a compact, lightweight, multi-sensor system, which we have integrated and demonstrated with various semiconductor and scintillator radiation detectors, shown in Figure 1. NG-LAMP consists of an onboard computer for all real-time data computation, contextual sensors for 3D mapping (including a LIDAR [7] and camera), a battery which enables about one hour of continuous data collection, and a 2x2 array of 1x1x2 in$^3$ CLLBC scintillators. The total system dimensions are 30x16x29 cm$^3$ (LxWxH) and it weighs 4.0 kg. NG-LAMP has been demonstrated in both a handheld configuration, as well as on a sUAS, results shown below. For measurements with the sUAS, we mounted the NG-LAMP system on a DJI Matrice 600 [8], shown in Figure 1 (right). NG-LAMP provides neutron and gamma-ray identification through pulse shape and pulse height discrimination. Pulse height discrimination is used as a matter of convenience for most results that follow, due to low fast neutron sensitivity in CLLBC [4-6].

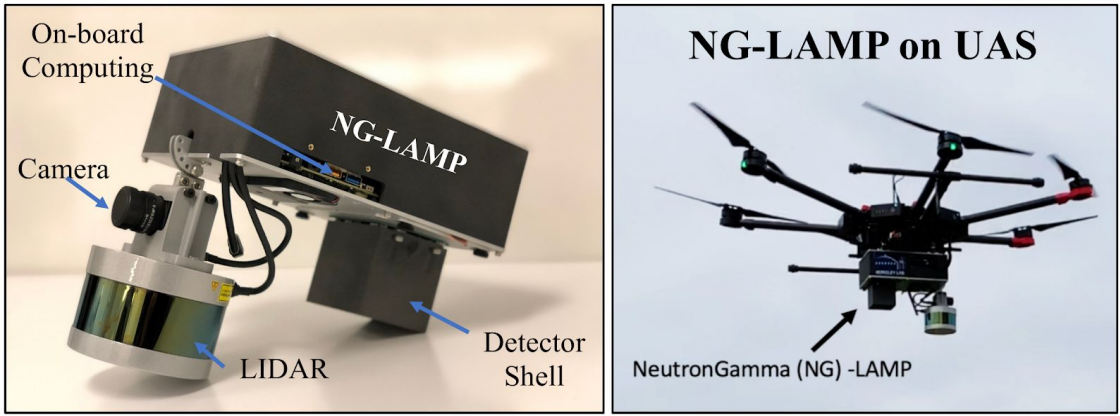

**Figure 1:** (Left) Picture of the NG-LAMP system, and (right) NG-LAMP mounted on sUAS (DJI Matrice 600).

NG-LAMP performs 3D SDF onboard in real-time. SDF requires a Simultaneous and Mapping (SLAM) engine, which is implemented in the LAMP software framework using Google Cartographer [9] to provide position and orientation estimates. Maximum likelihood expectation maximization (MLEM) [2] is performed in 3D for all poses, as illustrated in Figure 2, to reconstruct the most probable source distribution sources – be it a point source or a broad distribution. For each CLLBC module, a set of angular-dependent gamma-ray detection efficiencies are incorporated in the MLEM solution to provide an improved estimate of the source location.

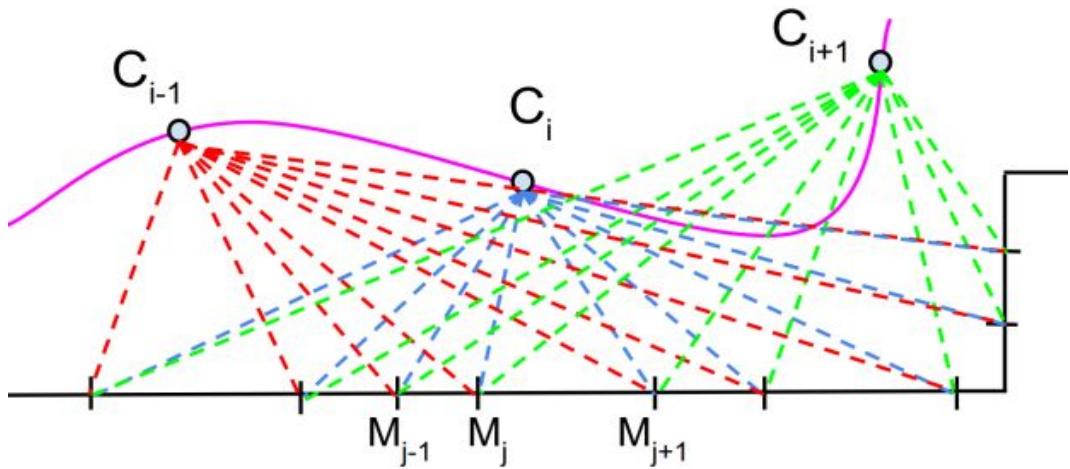

**Figure 2:** A schematic of the LAMP MLEM problem, where the Simultaneous Localization and Mapping solution already exists. Here the magenta curve represents the trajectory of the platform, $C_i$ represents the radiation data integrated into time bins (i subscript) associated with each position, and $M_j$ represents the 3D model derived from SLAM as points or 3D voxels.

**Results and Discussion**

*A. Simultaneous Real-Time Mapping of Gamma-ray and Shielded Fission Neutron Sources*

In the following demonstration, NG-LAMP was able to localize a neutron source in the presence of a much more active gamma-ray source. NG-LAMP participated in a concealed source demonstration at Bellows Air Force Base, HI in March 2019, during which a 500 μCi $^{60}$Co as well as Pu surrogate source with neutron emission rate equivalent to 1 SQ of Pu were placed in vehicles. NG-LAMP was deployed on a DJI Matrice 600 sUAS and flown around and over vehicles at a distance of about 1 m. The Pu surrogate source was shielded with 14 cm of lead in order to attenuate the gamma-ray signal from the source.

Figure 3 illustrates this measurement scenario (top-right and top-left), where the rental truck contains the 500 μCi $^{60}$Co source and the black Sports Utility Vehicle (SUV) contains the 1 SQ surrogate source. The two sources were separated by approximately 10m. The total measurement time for a single flight around this scene was eight minutes. Real-time data products were remotely streamed to the user in a top-down projection as NG-LAMP performed 3D SDF onboard the sUAS during the flight.

During this measurement, NG-LAMP successfully localized the $^{60}$Co source to the upper passenger's side corner of the rental truck's cargo compartment by using gamma-ray data for source reconstruction (Figure 3 bottom-left). Additionally, NG-LAMP produced a 3D neutron source localization of the shielded 1 SQ Pu surrogate source using only neutron data for reconstruction of the source distribution (Figure 3 bottom-right).

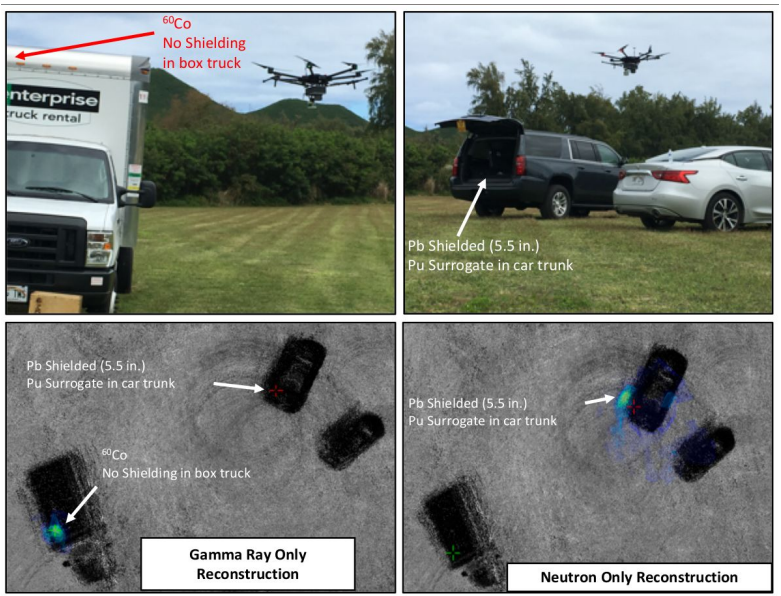

**Figure 3:** Simultaneous detection and localization of the neutron source (right) in the presence of a strong gamma-ray source (left). These 3D data products were produced from a single UAS flight of about eight minutes around a passenger car, a mid-size SUV and a rental truck. The $^{60}$Co (500 μCi) and mixed Pu surrogate source (~300 μCi $^{252}$Cf, $^{133}$Ba, $^{137}$Cs) were spatially separated by about 10 m. The neutron source was shielded with ~14 cm of lead, to reduce gamma-ray and x-ray intensity. The two figures were produced using the same reconstruction parameters, with only particle discrimination being applied to the figure on the right. The figure on the left includes both sources, however the neutron source is not visible due to the relative detected count rates from the two sources.

*B. Near Real-time Spectroscopic Localization of Gamma-Ray and Neutron Sources*

For this demonstration we setup a measurement with four spatially separated sources, to show NG-LAMP's ability to localize each source by its corresponding signature or by spectroscopic windowing for detected count rates. Figure 4 and 5 show spectroscopic data taken with NG-LAMP, during a two to three minute measurement, which involved walking through four rooms. These rooms contained three different 10 µCi gamma-ray sources ($^{137}$Cs, $^{60}$Co, $^{241}$Am/$^{133}$Ba) and a single 1 Ci PuBe neutron source. In this demonstration measurement we configured NG-LAMP to window on spectral photo-peaks for each of the sources to record counts in separate 3D reconstructions. The PuBe source was shielded with 14 cm of lead (Pb) and tungsten (W) to limit dose to personnel and to reduce gamma-ray emission. 2.5 cm of borated polyethylene was added around the lead and tungsten to reduce the neutron dose. The strength of the neutron source for this demonstration was substantial enough for neutrons to be detected through the adjoining 12-inch thick concrete wall. Therefore, the results of this measurement demonstrate our ability to simultaneously detect and localize the three individual gamma-ray sources and the α-n neutron source, rather than the minimum detectable activities for each. In Figure 4, we correctly localize the four source locations in near real-time, with the average rate of reconstruction updates at about 2 Hz on NG-LAMP for this scene. The survey of these four rooms represents a scan of the walls and bench space in the three 10x10 m$^2$ rooms during the three minute measurement.

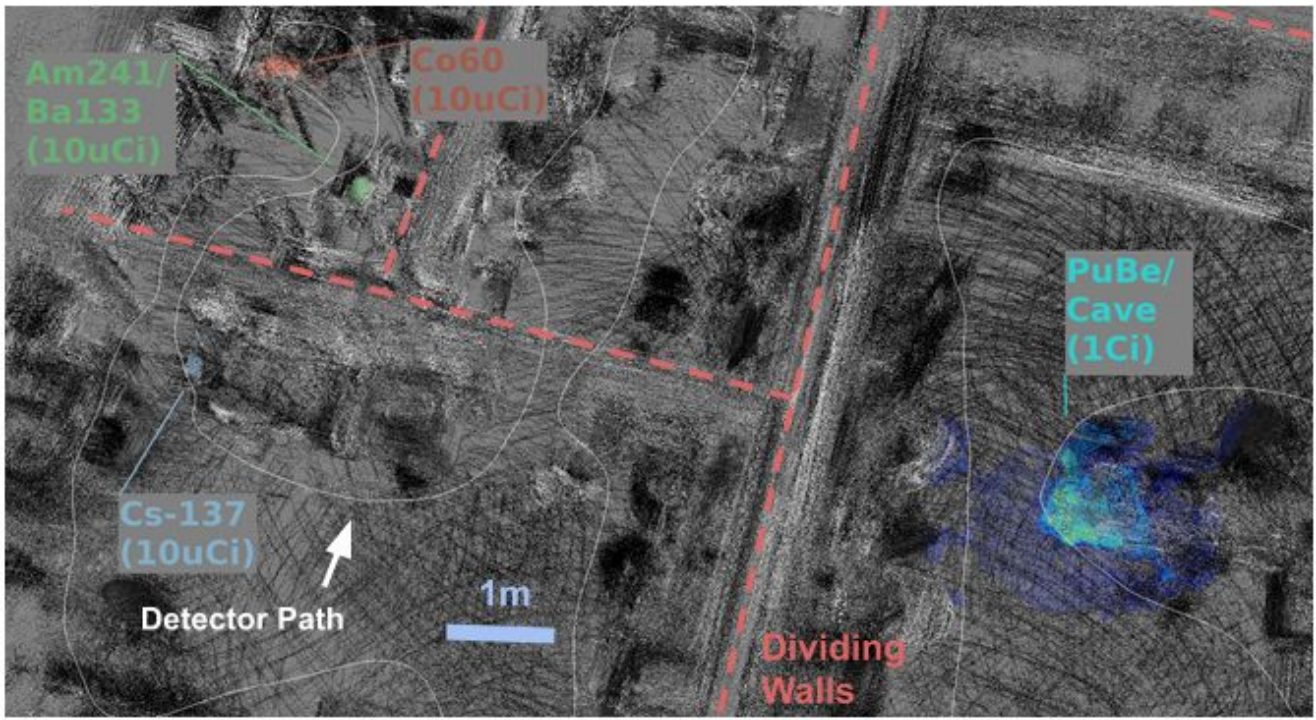

**Figure 4:** Simultaneous localization of three 10 µCi sources of $^{137}$Cs, $^{60}$Co, $^{241}$Am/$^{133}$Ba and 1 Ci PuBe source shielded by ~14 cm of Pb and W and 2.5 cm of borated polyethylene. The measurement for the four rooms (~10m x 10m each) took around three minutes. Gamma-ray sources were photo-peak windowed, as was the neutron source. Three iterations of MLEM were used to estimate source locations for photopeak or particle counts from each source.

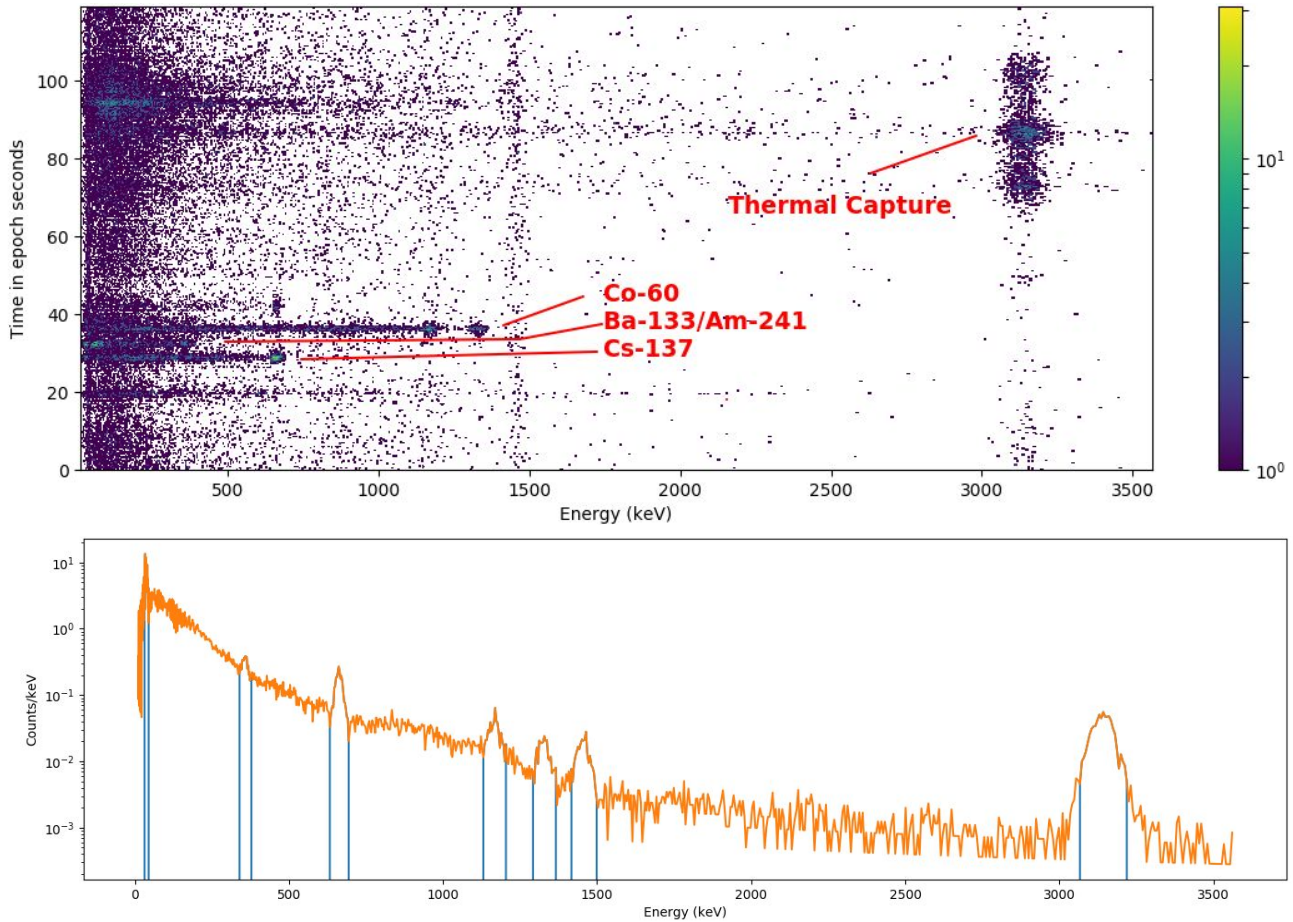

Figure 5: Energy spectra figures from data collected in Figure 4. (Top) A 2D plot of events in time vs energy detected. Here we clearly resolve the spectra correlated with $^{137}$Cs, $^{60}$Co, $^{241}$Am/$^{133}$Ba and thermal neutron capture. The colorbar is in counts/kev/100ms. (Bottom) Accumulated energy spectrum from the run.

*C. Simultaneous Real-Time Gamma-ray Background and Neutron Point Source Mapping with NG-LAMP of a Shielded 1 SQ Pu Surrogate*

Finally, NG-LAMP demonstrates the ability to localize a neutron source, which is heavily shielded by tungsten and lead. Additionally, with the same reconstruction parameters, NG-LAMP shows that the data is consistent with a near uniformly distributed gamma-ray background. In the measurement shown in Figure 6, we configured a 1 SQ Pu surrogate source, encased by a 14 cm-thick lead and tungsten cave, to produce a neutron source with minimal gamma-ray emission. This source and cave were placed in the back of a vehicle, where the gamma-ray emission from the source was consistent with background dose rates at the bumper of the car (~10 μR/hr). NG-LAMP was flown on a DJI Matrice 600 sUAS around the car at a distance of about 1 m for a total flight time of about three minutes. Data from the system, including a top-down projection of the real-time 3D reconstruction, were broadcasted to the control tablet and operator over a wireless network.

The 3D reconstruction that was produced in real-time from this measurement is shown in Figure 6 (bottom). This reconstruction uses data from this flight, for both gamma-ray and neutron data reconstructed in 3D, showing a localization of the neutron source in red and indicating no apparent point source from the gamma-ray data, the reconstruction of which is shown in blue. The gamma-ray reconstruction is consistent with a nearly uniform distribution of gamma-ray background. The simultaneous neutron/gamma SDF reconstruction capability is unique, in that it provides high resolution, 3D, gamma-ray and neutron mapping from the same platform, capable of flying on a sUAS, and in real-time.

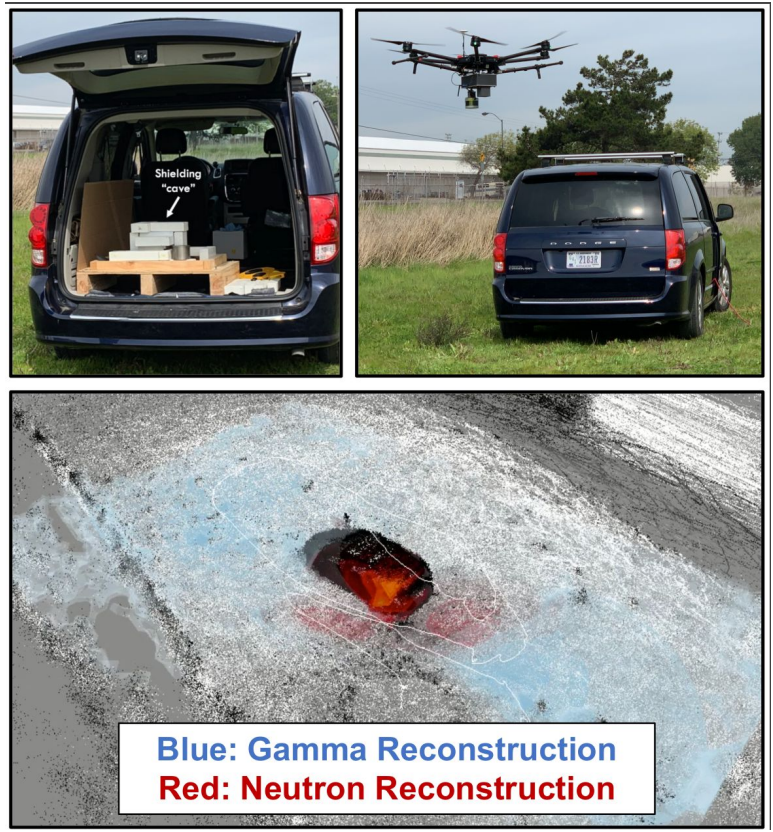

**Figure 6:** NG-LAMP reconstruction of simultaneous 3D real-time mapping of a shielded neutron source in the back of a vehicle. There are approximately 14 cm of lead and tungsten for shielding of a 1 SQ Pu surrogate source, which reduces the dose rate from gamma-ray emission, measured at the bumper, to about 10 μR/hr. Thus the source emission is considered to be purely neutrons. The reconstruction is consistent with the uniform background gamma-ray distribution, shown in blue, and the neutron distribution shown in red.

**Conclusion**

Utilizing a neutron and high-resolution gamma-ray detection platform, NG-LAMP, we have demonstrated the ability to simultaneously localize gamma and neutron sources in real-time. These capabilities were demonstrated in several scenarios: in the presence of strong gamma-ray sources and where gamma-ray signatures are concealed by substantial shielding. We have demonstrated this capability with both fission and α-n neutron sources. We have also demonstrated the spectroscopic capability of this system with photo-peak windowing techniques in near real-time. Additionally, we have demonstrated the ability to produce localized distributions for point sources and near uniform distributions. Finally, we have demonstrated NG-LAMP in both handheld and sUAS-mounted configurations. We did not alter gamma-ray or neutron reconstruction parameters for MLEM, except for gamma-ray and neutron energy cuts, as we have specified. Switching between different deployment configurations, including sUAS, ground vehicle-mounted, or handheld, can be done through simple and fast (e.g. less than five minutes) mechanical mount changes for the desired mode of operation. This contribution increases the utility of nuclear radiation mapping, extending the concepts of 3D SDF to neutron sources and for the first time, localizing neutron sources in 3D and real-time with a four monolithic scintillator volumes in measurements lasting <10 minutes.

 Acknowledgements

This material is based upon work supported by the Defense Threat Reduction Agency under IAA numbers 10027-28022 and 10027-23334. This support does not constitute an express or implied endorsement on the part of the United States Government.